\documentclass{IEEEcsmag}

\usepackage[colorlinks,urlcolor=blue,linkcolor=blue,citecolor=blue]{hyperref}
\expandafter\def\expandafter\UrlBreaks\expandafter{\UrlBreaks\do\/\do\*\do\-\do\~\do\'\do\"\do\-}
\usepackage{upmath,color}
\usepackage{xspace}

\jvol{XX}
\jnum{XX}
\paper{8}
\jmonth{Apr}
\jname{Publication Name}
\jtitle{Publication Title}
\pubyear{2025}

\newcommand{\etal}[1]{et~al.}

\begin{document}

% \sptitle{Article Type: Description}
\sptitle{Preprint}
% \sptitle{Article Type: Feature Article}

\title{Don't Trust the Label: License Laundering in AI Supply Chains}

\author{James Jewitt}
\affil{Queen's University}

\author{Hao Li}
\affil{Queen's University}

\author{Gopi Krishnan Rajbahadur}
\affil{Queen's University}

\author{Bram Adams}
\affil{Queen's University}

\author{Ahmed E. Hassan}
\affil{Queen's University}

% \markboth{THEME/FEATURE/DEPARTMENT}{THEME/FEATURE/DEPARTMENT}
% \markboth{PREPRINT}{PREPRINT}
\markboth{PREPRINT}{PREPRINT}
% \markboth{FEATURE}{FEATURE}

% Three actionable insights (IEEE Software required deliverable; draft, finalize at submission):
% For practitioners: a model's license label is not evidence of upstream rights. In the 232,270 supply chains we traced, 62.3% pass through at least one artifact with no declared license. Verify the declared training datasets' licenses before integrating a model, and treat non-disclosure as a risk in itself.

% For model publishers: declare your training datasets and their license categories in Hugging Face's existing datasets field. Only 7.1% of models declare their training data today, and in the chains we could trace, most laundering happens at the dataset-to-model hop. Closing that gap requires documentation practice, not new infrastructure.

% For platform owners: surface upstream license categories next to downstream labels and flag conflicts, such as a CC BY-SA dataset training an MIT-labeled model. No software composition analysis tool follows the chain across Hugging Face and GitHub, so the platforms themselves are best positioned to close this gap.
\begin{abstract}

\looseness-1
AI artifacts move through a multi-platform supply chain, spanning datasets and models on Hugging Face and applications on GitHub. While each artifact carries a license whose obligations should propagate through redistribution, no study has yet measured whether those obligations survive the chain or are stripped and replaced as artifacts move downstream. We trace 232,270 dataset$\rightarrow$model$\rightarrow$application chains and quantify two forms of \textit{license laundering}: when artifacts with no declared license acquire definitive labels downstream, and when one declared license category replaces another during redistribution. We find that 62.3\% of chains pass through at least one artifact with no declared license (concentrated in a small set of foundational datasets), and that every obligation-bearing license category falls below 7\% end-to-end survival while the Permissive category reaches 95.1\%. Based on these findings, we provide actionable recommendations for practitioners, model publishers, rights holders, and platform owners.
\end{abstract}

\maketitle

\chapteri{S}entence-transformers/all-MiniLM-L6-v2, labeled Apache-2.0 on Hugging Face, is integrated by 1,656 applications on GitHub. While a practitioner reading that license label likely assumes the model is freely reusable, of its 21 declared training datasets, (1) nine carry no declared license, which leaves all rights reserved to the dataset's creators, and (2) three carry Sharealike terms that require any derivative to preserve the same obligations. The Apache-2.0 license label does not reflect those upstream rights, i.e., this is an example of \textit{license laundering}: the stripping or replacement of the legal rights and obligations attached to an asset as it moves through an AI supply chain. 
 
We focus on two forms of license laundering: \textit{Unknown laundering} occurs when artifacts with no declared rights acquire definitive licenses downstream, implying legal certainty where none was established, as with the nine no-license datasets above. \textit{Category laundering} occurs when one declared license category is replaced with another during redistribution, as with the three Sharealike datasets whose obligations vanish under the model's Apache-2.0 label. Both types of laundering introduce legal uncertainty: an upstream rights holder can assert copyright at any time, forcing a takedown, mandatory relicensing, or financial liability on every application that trusted the laundered license.

Books3 illustrates both forms of laundering and their cost. The dataset packaged 196,640 books from the pirate library Bibliotik, whose rights holders granted no downstream rights, yet circulated under an MIT label, i.e., \textit{Unknown laundering}. The label then propagated. Models trained on Books3, from Anthropic's Claude to LLaMA and BloombergGPT, shipped under their own license terms that replaced the dataset's label, i.e., \textit{Category laundering}, leaving practitioners no signal that the underlying rights were never established. In 2023, the Danish Rights Alliance forced Books3's withdrawal to stop AI from becoming ``an opportunity to launder copyright infringement''~\cite{books3-takedown}. The cost surfaced in \textit{Bartz v.\ Anthropic}: a federal court held that training on the books was fair use, but that acquiring and retaining pirated copies was not; Anthropic settled for \$1.5B, approved in July 2026~\cite{bartz-anthropic}. Because liability followed provenance, publishers that ingested the same sources face the same claims, and downstream applications inherit takedown and relicensing risk. Understanding, quantifying, and preventing laundering are therefore essential, and decades of software-license enforcement show how routinely such risks materialize: the Court of Milan held that BSD-4-Clause non-compliance retroactively strips the license, reclassifying commercial distribution as infringement~\cite{milan7112-2023}, and Terraform's relicensing drove more than 140 companies to the OpenTofu fork within six weeks~\cite{opentofu-fork}.

No existing tool is able to automatically audit the licenses along an AI supply chain for laundering occurrences. The software industry has created software composition analysis (SCA) tools such as FOSSology and Black Duck that trace declared software licenses through package-manager dependency graphs, flag incompatible combinations, and generate Software Bills of Material (BOMs) for compliance audits~\cite{ombredanne2020sca}. However, these tools cannot be readily used on AI supply chains, as the latter break the assumptions existing software license auditing tools rely on. A model's training-data dependencies are declared not in lockfiles but in metadata tags on Hugging Face, and an application's model dependencies can only be recovered through code search across GitHub. Often no license is declared at all: 65.0\% of models and 74.7\% of datasets on Hugging Face carry no recognizable license tag~\cite{stalnaker}. AI chains also cross platforms: a Hugging Face dataset trains a Hugging Face model that a GitHub application then integrates, a path no SCA tool follows end-to-end. As a result, the warnings that SCA tools routinely raise for software, e.g., that a copyleft dependency requires the integrating project to share its source under the same terms, or that two licenses in a dependency graph conflict, are never produced for AI artifacts. In other words, two decades of open-source software (OSS) compliance infrastructure has no AI equivalent. 

Building that tooling first requires knowing how often laundering happens. Prior audits quantify how poor Hugging Face's license labels are (over 96\% of permissively-labeled artifacts lack the full license text their declared licenses require~\cite{jewitt2026permissivewashing}), but poor labels are not the same as lost obligations. What no prior work has measured is whether the obligation-bearing license categories (those that constrain how an artifact may be used or reused) survive the full dataset-to-model-to-application chain. Without that measurement, we cannot tell whether laundering is rare and isolated or routine and systemic, whether it concentrates in a few foundational artifacts or spreads across the ecosystem, or whether different categories are laundered at different rates.

This paper provides the first empirical measurement of how license categories change across the full AI supply chain. We trace 232,270 dataset$\rightarrow$model$\rightarrow$application chains spanning Hugging Face and GitHub, measure both forms of laundering, and identify the recurring patterns behind them. Our contributions are:
\begin{itemize}
    \item We quantify Unknown laundering across 232,270 chains, finding that \textbf{62.3\% pass through at least one artifact with no declared license}, and show that the pattern concentrates in a small number of foundational datasets.
    \item We measure Category laundering at each transition (a dataset$\rightarrow$model, model$\rightarrow$application or dataset$\rightarrow$application), finding that \textbf{every obligation-bearing license category falls below 7\% end-to-end survival, while the Permissive category reaches 95.1\%}.
    \item We translate these patterns into concrete risks and actions for practitioners, model publishers, rights holders, and platform owners building license-tracing infrastructure.
\end{itemize}

\section{AI Supply Chain Composition}

To measure how license categories change across AI artifacts, we construct dataset$\rightarrow$model$\rightarrow$application chains spanning Hugging Face and GitHub, then categorize each artifact's licenses into seven categories by the obligations they impose. Full collection details are in our replication package.\footnote{\url{https://github.com/SAILResearch/LicenseLaundering/}}

\subsection{Supply Chain Construction} 

Following a methodology similar to our prior work~\cite{jewitt2026permissivewashing}, we link Hugging Face datasets and models to the GitHub applications that use them, then extract each artifact's declared license. We link datasets to models through Hugging Face's dataset dependency metadata, identify GitHub applications that invoke those models via code search and abstract syntax tree (AST), and use ScanCode to extract the license label declared in each repository. This initial collection contains 3,198 datasets, 6,218 models, and 26,302 applications, forming 294,012 candidate dataset$\rightarrow$model$\rightarrow$application chains. 

We then apply two filters on the candidate chains. The first filter removes models that serve only as base models with no application directly invoking them, ensuring that every chain passes through exactly two transitions, which leaves 264,431 chains. The second filter excludes any chain in which an artifact carries a license string that does not map to our classification (described next), removing a further 32,161 chains. After both filters, we are left with 232,270 chains spanning 3,120 datasets, 5,556 models, and 24,076 applications. Of these, 144,631 (62.3\%) pass through at least one artifact with an Unknown license, and 87,639 (37.7\%) carry a Known category at all three artifacts. 

\subsection{License Categorization}

Each AI artifact on Hugging Face~\cite{hf-model-cards} or GitHub~\cite{github-licensing} carries one or more license strings in its metadata. Across all artifacts, we observe 765 unique strings. Tracking individual strings is too granular for measuring how obligations change across a supply chain, since licenses imposing the same type of restriction (e.g., MIT and BSD-3-Clause) belong together~\cite{stalnaker}. For this reason, we group strings into seven categories based on the obligations they impose, following classification guides from the Free Software Foundation (FSF)~\cite{fsf-licenses}, Creative Commons (CC)~\cite{cc-compatible}, and Stalnaker et al.~\cite{stalnaker}:

\begin{itemize}
    \item \textbf{Permissive} (e.g., MIT, Apache-2.0, BSD-3-Clause)
    \item \textbf{Copyleft} (e.g., GPL, AGPL)
    \item \textbf{Sharealike} (e.g., CC BY-SA, LGPL, MPL)
    \item \textbf{ML License} (e.g., OpenRAIL, Llama Community License)
    \item \textbf{CC-Restrictive} (grouping NonCommercial, NC-SA, NC-ND, and NoDerivatives, which share the practical constraint of restricting downstream use)
    \item \textbf{Public Domain} (e.g., CC0, Unlicense)
    \item \textbf{Unknown}
\end{itemize}

 \noindent The Unknown category combines all license strings that represent absent or undeclared rights (e.g., empty fields, ``Unknown,'' ``Other'').

We manually assign each license string to one of our seven categories when it matches a license recognized by Hugging Face or GitHub, or an entry in the FSF and CC classification guides, giving 270 classified strings. The remaining 495 of the 765 unique strings match no entry in the classification, and we label them as unidentified. An artifact can carry multiple license categories simultaneously (e.g., a dataset licensed under both Permissive and Sharealike terms).

\section{Unknown Laundering}

\begin{figure*}[t]
	\centering
    \includegraphics[width=0.89\textwidth]{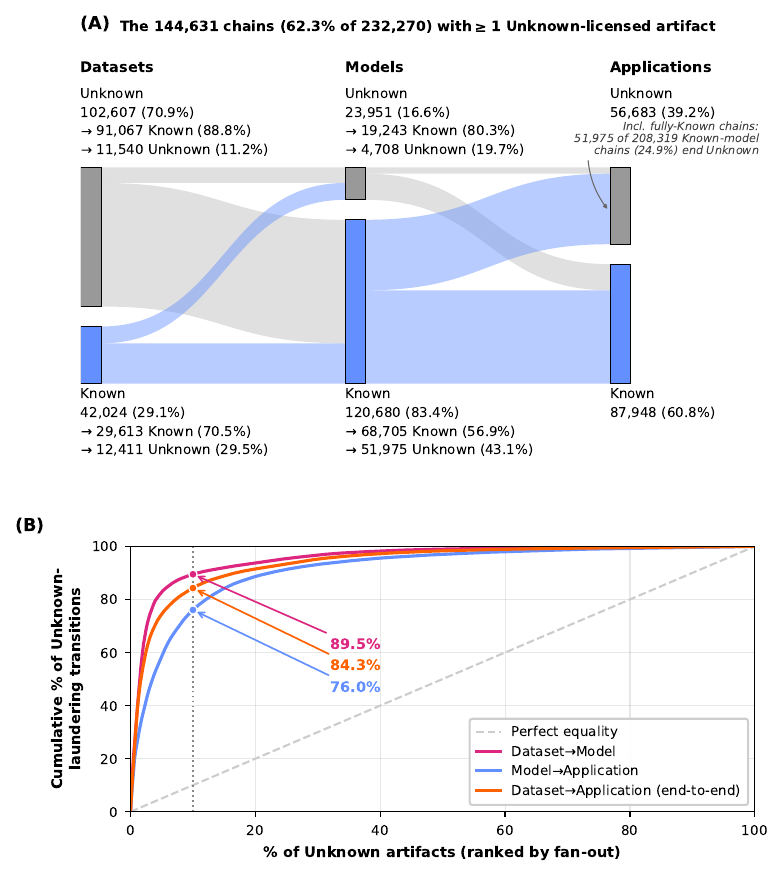}
	\caption{\textbf{(A):} Sankey diagram of the 144,631 chains that pass through at least one Unknown-licensed artifact; the 87,639 fully-Known chains are analyzed in Figure~\ref{fig:figure_2}. \textbf{(B):} Lorenz curves: Unknown-laundering transitions concentrate in the top 10\% of Unknown artifacts.}
	\label{fig:figure_1}
\end{figure*}

Unknown laundering has direct legal consequences: when an artifact carries no declared license, the rights holder retains full copyright by default~\cite{github-licensing}, and any downstream label that appears settled rests on rights that were never established.

The model facebook/convnext-base-224, for instance, trains on ImageNet-1K yet ships under an Apache-2.0 label, even though ImageNet-1K declares no license and so grants none of the permissions that label implies. We measure the prevalence of this pattern, the concentration of upstream Unknown artifacts that drive the pattern,  and its inverse pattern (where declared rights are dropped on the way downstream).

We classify every dataset$\rightarrow$model and model$\rightarrow$application transition into one of four types: Unknown$\rightarrow$Known (Unknown laundering), Known$\rightarrow$Unknown (inverse of Unknown laundering), Unknown$\rightarrow$Unknown, or Known$\rightarrow$Known (analyzed in Section~\nameref{sec:categoryl}). When an artifact carries multiple license categories (e.g., Permissive and Sharealike), we classify it as Known if any category is Known. The Unknown laundering rate is the proportion of Unknown-upstream transitions whose downstream artifact carries a Known license, measured at the dataset$\rightarrow$model, model$\rightarrow$application transitions, and end-to-end (Figure~\ref{fig:figure_1}(A)). To assess whether Unknown laundering is concentrated or dispersed, we rank Unknown artifacts by the number of laundering transitions they generate and plot Lorenz concentration curves: the further the curve bows from the diagonal, the more concentrated the transitions are in a small number of artifacts (Figure~\ref{fig:figure_1}(B)).

\textbf{62.3\% of supply chains pass through at least one artifact with an Unknown license.} Of the 232,270 supply chains in our dataset, 144,631 contain at least one such artifact (Figure~\ref{fig:figure_1} (A)). An Unknown license anywhere in the chain means the downstream label does not rest on verified upstream rights, regardless of how settled it appears. This pattern is bidirectional. In the laundering direction, Unknown artifacts acquire definitive licenses at both hops: 88.8\% of chains with an Unknown dataset reach a Known model, and 80.3\% of chains with an Unknown model reach a Known application (Figure~\ref{fig:figure_1}(A)). In the other direction, declared rights vanish: 24.9\% of all chains with a Known-licensed model end at an application with no declared license, erasing obligations that should propagate. A license backed by verified upstream rights looks identical to one with no basis at all.

\textbf{Unknown laundering concentrates in a few foundational datasets: the top 10\% of Unknown datasets account for 89.5\% of dataset$\rightarrow$model laundering transitions.} Figure~\ref{fig:figure_1}(B) shows the Lorenz curves for all three transitions. Concentration is highest at the dataset$\rightarrow$model transition and remains high end-to-end (84.3\%) and at model$\rightarrow$application (76.0\%). ImageNet-1K, a widely used computer-vision training dataset, carries an Unknown license. It fans out to 243 models, 208 of which are Known, reaching 1,709 applications. The Pile and BookCorpus follow the same pattern, reaching 2,139 and 1,695 applications respectively. The MiniLM example from the introduction is a downstream symptom of this dynamic: a small number of foundational Unknown datasets generate most of the laundering throughout the chain.

\textbf{Even when upstream rights are clearly declared, downstream artifacts still drop them: one in four chains with a Known-licensed model (24.9\%) ends at an application with no declared license.} The inverse of Unknown laundering, i.e., declared obligations transitioning to Unknown, is common in its own right (Figure~\ref{fig:figure_1}(A)): of the 144,631 chains involving an Unknown artifact, 56,683 (39.2\%) end at an Unknown application. The drop concentrates at the model$\rightarrow$application transition. Developers may not recognize that integrating a model creates a derivative work, or they may assume that an application's own license overrides upstream terms. AK391/ai-gradio integrates upstage/SOLAR-10.7B-Instruct-v1.0, a model released under CC-BY-NC-4.0 prohibiting commercial use, but the application carries an Unknown license. Nothing signals the upstream restriction to a developer who uses the application commercially. Cross-platform tooling that warned them when an application inherits a model's restriction, such as a non-commercial clause, would close this gap, but none exists.

\section{Category Laundering}
\label{sec:categoryl}

\begin{figure*}[!tbh]
	\centering
	\includegraphics[width=\textwidth]{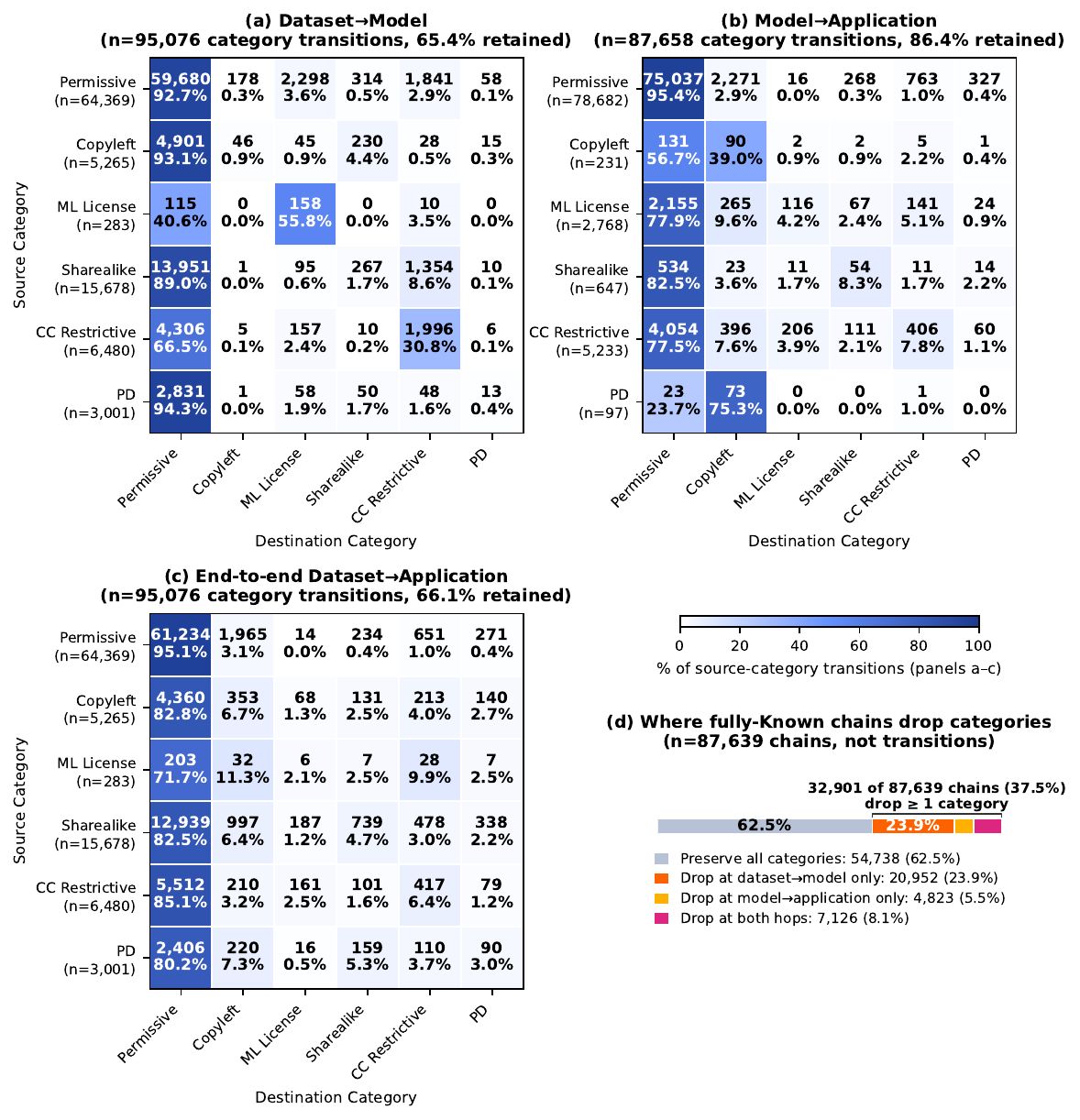}
	\caption{\textbf{(A)} Dataset$\rightarrow$model, \textbf{(B)} model$\rightarrow$application, and \textbf{(C)} end-to-end dataset$\rightarrow$application transition matrices showing how upstream license categories map to downstream categories; diagonal cells indicate retention. The panels count category transitions rather than chains: the 87,639 fully-Known chains yield 95,076 dataset$\rightarrow$model and 87,658 model$\rightarrow$application category transitions, since artifacts with multiple license categories contribute one transition per category. \textbf{(D)} Where the 32,901 chains (37.5\%) that drop at least one category do so.}
	\label{fig:figure_2}
\end{figure*}

A fully-Known chain avoids Unknown laundering, but its declared license categories can still change in transit. Category laundering occurs when a downstream artifact carries a different license category than the upstream artifact it derives from, replacing the creator's original choice. We measure it on the 87,639 chains where all three artifacts carry at least one Known license category. We analyze each license category separately, and a dataset with licenses in more than one category is counted in each category's analysis. This yields 95,076 category observations across datasets and 87,658 across models. We then build transition matrices at the dataset$\rightarrow$model and model$\rightarrow$application steps (Figure~\ref{fig:figure_2}(A,B)) and an end-to-end matrix (Figure~\ref{fig:figure_2}(C)). End-to-end retention can exceed single-hop retention because a category dropped at the model may reappear at the application, e.g., when an application's own license coincidentally matches the dataset's category. We classify a transition as laundered if any upstream category is absent from the downstream artifact's category set. The survival rate is the fraction of chains where all obligation-bearing upstream license categories propagate to the application.

\textbf{37.5\% of fully-Known supply chains contain at least one laundered transition.} Of the 87,639 chains where every artifact carries a Known license, 32,901 drop at least one category somewhere along the chain (Figure~\ref{fig:figure_2}(D)). A chain can drop a category without any Unknown license. By definition, any Sharealike dataset whose model is labeled only Permissive has dropped its Sharealike category, though both artifacts are Known. Most laundering happens at the dataset$\rightarrow$model transition, where only 65.4\% of category transitions retain their category, versus 86.4\% at model$\rightarrow$application (panel titles, Figure~\ref{fig:figure_2}(A,B)). Figure~\ref{fig:figure_2}(D) locates the drops per chain: 23.9\% of fully-Known chains drop a category only at dataset$\rightarrow$model, 5.5\% only at model$\rightarrow$application, and 8.1\% at both hops. Verifying only the model and application licenses misses the obligations a dataset carried but a downstream artifact dropped.

\textbf{Permissive licenses survive end-to-end at 95.1\% (Permissive diagonal, Figure~\ref{fig:figure_2}(C)), while every obligation-bearing license category falls below 7\%.} Sharealike is the upstream category in 15,678 chains, but survives to the application in only 739 of them (4.7\%; Sharealike diagonal, Figure~\ref{fig:figure_2}(C)), roughly a twentieth of Permissive's rate. When an obligation-bearing license category is lost, the downstream license almost always becomes Permissive rather than another category. For dataset$\rightarrow$model, 93.1\% of Copyleft transitions land on Permissive (Copyleft row, Figure~\ref{fig:figure_2}(A)). For model$\rightarrow$application, 77.9\% of ML License transitions land on Permissive (ML License row, Figure~\ref{fig:figure_2}(B)).

Model and application developers know how to ship Permissive software, so a likely explanation for the collapse toward Permissive is that when these developers hit an upstream category they cannot satisfy, they label their own artifact Permissive or drop the license entirely. For example, the Wikipedia dataset, which carries both Copyleft and Sharealike, originates 4,849 chains, yet only 86 preserve both end-to-end (1.8\%). The wukong100m dataset, released under CC-Restrictive terms prohibiting commercial use and derivatives, has 89 downstream models, none of which carry CC-Restrictive forward. In both cases, the restriction the rights holder chose barely survives downstream, if at all. 

\textbf{The ML License and Copyleft categories mirror each other: ML License survives dataset$\rightarrow$model (55.8\%) but not model$\rightarrow$application (4.2\%), while Copyleft does the opposite (0.9\% then 39.0\%).} Models adopt ML License terms from training data at a moderate rate. Applications then strip them (ML License diagonals, 55.8\% in Figure~\ref{fig:figure_2}(A) versus 4.2\% in (B)). For instance, StarCoder, a widely used 15.5B-parameter code-generation model,\footnote{\url{https://huggingface.co/bigcode/starcoder}} carries an ML License, but 314 of its 321 Known downstream applications drop it. ML License terms prohibit downstream uses such as surveillance or medical decision-making, yet for the 314 applications that dropped the label, no record of those prohibitions survives. The licenses provide no operational mechanism for enforcing these prohibitions in software, and behavioral-use licensing has no precedent in conventional software composition tools. Copyleft, by contrast, is native to the software ecosystem where applications live and has decades of tooling support behind it. 

\section{Threats to Validity}

\textbf{Internal Validity.} We track license labels, not the full legal text of the licenses. For datasets and models, these labels come from self-reported metadata on Hugging Face. For applications, we use the license label identified by ScanCode from GitHub repositories. In all cases, if the label does not reflect the publisher's intent, our analysis captures the label as practitioners encounter it, not the underlying legal status. License strings with vague or ambiguous metadata (e.g., ``other'') are classified as Unknown alongside absent licenses. If some of these strings represent intentional license choices that we cannot resolve, the Unknown count may be overstated. Stalnaker et al.~\cite{stalnaker} handle such strings the same way, treating Hugging Face's ``other'' and ``unknown'' labels as unresolvable categories rather than mapping them to specific licenses, since retrieving and evaluating these license terms is difficult and time-consuming.

\textbf{External Validity.} Our collected data includes only the models that declare their training data via metadata, which amount to 7.1\% of Hugging Face models~\cite{jewitt2026permissivewashing}. Models that do not disclose training data and applications that do not appear in GitHub code search are outside our scope. Prior audits of the Hugging Face supply chain face the same boundary: Stalnaker et al.~\cite{stalnaker} build their supply-chain graph from declared metadata alone and note that it necessarily captures a subset of the full chain. The impact of this selection bias is unclear: models that disclose their training data may follow better or worse license-compliance practices than those that do not. 

Our sample is drawn from the most popular models and datasets, so it captures the artifacts with the widest downstream reach, including high-fan-out foundational datasets such as ImageNet-1K, The Pile, and BookCorpus. Our findings therefore characterize the popular, disclosed portion of the supply chain rather than its long tail. We also exclude chains containing license strings that do not map to our classification framework (495 of 765 unique strings, predominantly custom or ambiguous). The excluded chains represent only 12.2\% of the pre-filter chains. Every standard obligation-bearing license (Copyleft, Sharealike, ML License, CC-Restrictive) falls within the 270 strings we recognize, so the excluded strings do not affect the categories whose survival we measure.

\section{Implications}

\textbf{For practitioners: verify upstream before integrating.} The model's license alone is an unreliable signal. Unknown upstream artifacts produce definitive downstream labels in 91,067 of 232,270 chains (Figure~\ref{fig:figure_1}(A)), and obligation-bearing upstream categories collapse into Permissive downstream labels (Figure~\ref{fig:figure_2}(C)). Before integrating an artifact, practitioners should follow the model's declared training datasets on Hugging Face, classify the upstream licenses against the same categories the model claims, and treat any conflict or absence as an integration risk. When the model does not declare its training data, treat that absence as a risk signal in itself. 

\textbf{For model publishers: declare upstream dataset licenses, not just the model's own.} Only 7.1\% of Hugging Face models declare their training datasets via metadata~\cite{jewitt2026permissivewashing}. Even within that 7.1\%, 37.5\% of fully-Known chains contain at least one laundered transition. Publishers who omit upstream data pass undisclosed legal risk to every downstream practitioner. Hugging Face model cards already support a \texttt{datasets} field~\cite{hf-model-cards}; publishers should populate it and record the license category of each training dataset. This requires no new infrastructure, only a change in documentation practice.

\textbf{For rights holders: a restrictive license is not enforcement.} Permissive licenses survive end-to-end at 95.1\% while every obligation-bearing category falls below 7\% (Figure~\ref{fig:figure_2}(C)). ML-specific licenses, designed to prohibit uses such as surveillance, survive only 4.2\% of model$\rightarrow$application transitions (ML License diagonal, Figure~\ref{fig:figure_2}(B)). A rights holder who needs downstream compliance cannot rely on label propagation alone and should pair the license with enforcement mechanisms outside metadata: contractual terms or gated access. 

\textbf{For platform owners and tool builders: build the cross-platform tracer that does not yet exist.} Existing SCA tools (ScanCode, FOSSology, Black Duck) operate on single-platform, package-manager dependencies. AI chains span Hugging Face and GitHub, mix three legal paradigms, and declare dependencies through metadata tags. No tool traces cross-platform license propagation or flags cross-paradigm incompatibilities such as a CC BY-SA dataset training an MIT-labeled model. SPDX 3.0~\cite{spdx3} and CycloneDX~\cite{cyclonedx} both offer BOM profiles for AI artifacts that can represent the full chain, but nothing populates them automatically. A recent systematic review confirms that mitigation remains the least developed area of license compliance research~\cite{li2025oss}, and AI's cross-paradigm structure makes the gap more acute. Concretely: Hugging Face and GitHub should auto-surface upstream license categories alongside downstream labels and warn when categories conflict; researchers should build tools that auto-generate AI BOMs from existing metadata, connecting the existing standards to the supply chains they are designed to describe. Until then, manual upstream verification is the only safeguard.

\def\refname{REFERENCES}

\vspace*{-8pt}

\begin{IEEEbiography}{James Jewitt}{\,} is a Ph.D. student at Queen’s University, Kingston, ON K7L 3N6, Canada. Their research interests include AI supply chains, software and dataset licensing, and data provenance. Contact them at james.jewitt@queensu.ca.\vspace*{8pt}
\end{IEEEbiography}

\begin{IEEEbiography}{Hao Li}{\,} is a postdoctoral researcher at Queen’s University, Kingston, ON K7L 3N6, Canada. His research interests include software engineering for AI, AI for software engineering, and software package ecosystems. Li received his Ph.D. in Software Engineering and Intelligent Systems from the University of Alberta. Contact him at hao.li@queensu.ca.\vspace*{8pt}
\end{IEEEbiography}

\begin{IEEEbiography}{Gopi Krishnan Rajbahadur}
is a Principal Researcher at Huawei’s Centre for Software Excellence in Canada, leading data and post-training for the Pangu Foundation Model’s software engineering capabilities. He is also a Research Associate at Queen’s University’s Software Engineering and Analysis Lab, where he studies AI supply chains and trustworthy AI-agent ecosystems. His research spans software engineering for AI-powered systems and AI dataset governance and compliance. Contact him at grajbahadur@acm.org.
\end{IEEEbiography}

\begin{IEEEbiography}{Bram Adams} {\,} is a full professor at Queen’s University, Kingston, ON K7L 3N6, Canada. His research interests include software release engineering (pre- and post-AI) and mining software repositories. He is a Senior Member of IEEE. Contact him at bram.adams@queensu.ca.\vspace*{8pt}
\end{IEEEbiography}

\begin{IEEEbiography}{Ahmed E. Hassan} {\,} is the Natural Sciences and Engineering Research Council of Canada/Research in Motion Industrial Research chair in Software Engineering for Ultra Large Scale systems at Queen’s University, Kingston, ON K7L 3N6, Canada. He is a Fellow of the IEEE. Contact him at ahmed@cs.queensu.ca.
\end{IEEEbiography}


\begin{thebibliography}{1}

\bibitem{books3-takedown}
Rights Alliance Denmark, ``Rights Alliance Removes the Illegal Books3 Dataset Used to Train Artificial Intelligence,'' 2023. [Online]. Available: \url{https://rettighedsalliancen.com/rights-alliance-removes-the-illegal-books3-dataset-used-to-train-artificial-intelligence/}

\bibitem{bartz-anthropic}
\textit{Bartz et al. v. Anthropic PBC}, No. 3:24-cv-05417 (N.D. Cal.). Summary judgment order, Jun. 23, 2025; settlement granted final approval Jul. 20, 2026. See B. Brittain, ``US Judge Approves Anthropic's \$1.5 Billion Settlement of Copyright Lawsuit,'' Reuters, Jul. 20, 2026. [Online]. Available: \url{https://finance.yahoo.com/technology/ai/articles/us-judge-approves-anthropics-1-204851948.html}

\bibitem{milan7112-2023}
Tribunale di Milano, Sez. Imprese, Sentenza n. 7112/2023 (\textit{Gestionale Open}), deposited 18 September 2023. Commentary at \url{https://www.canellacamaiora.com/open-source-licenses-and-copyright-the-milan-court-on-license-violation/}.

\bibitem{opentofu-fork}
The Linux Foundation, ``Linux Foundation Launches OpenTofu: A New Open Source Alternative to Terraform,'' Press Release, Sept. 20, 2023. [Online]. Available: \url{https://www.linuxfoundation.org/press/announcing-opentofu}

\bibitem{ombredanne2020sca}
P. Ombredanne, ``Free and Open Source Software License Compliance: Tools for Software Composition Analysis,'' \textit{IEEE Computer}, vol. 53, no. 10, pp. 105--109, Oct. 2020.

\bibitem{stalnaker}
T. Stalnaker, N. Wintersgill, O. Chaparro, L. A. Heymann, M. Di Penta, D. M. German, and D. Poshyvanyk, ``The ML Supply Chain in the Era of Software 2.0: Lessons Learned from Hugging Face,'' \textit{arXiv preprint arXiv:2502.04484}, 2025.

\bibitem{jewitt2026permissivewashing}
J. Jewitt, G. K. Rajbahadur, H. Li, B. Adams, and A. E. Hassan, ``Permissive-Washing in the Open AI Supply Chain: A Large-Scale Audit of License Integrity,'' in \textit{Proc. 32nd ACM SIGKDD Conf. Knowledge Discovery and Data Mining V.2 (KDD '26)}, Jeju Island, Republic of Korea, 2026, 12 pages, doi: 10.1145/3770855.3818130. arXiv:2602.08816.

\bibitem{hf-model-cards}
Hugging Face, ``Create and share Model Cards,'' 2023. [Online]. Available: \url{https://huggingface.co/docs/huggingface_hub/en/guides/model-cards}

\bibitem{github-licensing}
GitHub, ``Licensing a repository,'' 2024. [Online]. Available: \url{https://docs.github.com/en/repositories/managing-your-repositorys-settings-and-features/customizing-your-repository/licensing-a-repository}

\bibitem{fsf-licenses}
Free Software Foundation, ``Various Licenses and Comments about Them,'' 2026. [Online]. Available: \url{https://www.gnu.org/licenses/license-list.html}

\bibitem{cc-compatible}
Creative Commons, ``Compatible Licenses,'' 2026. [Online]. Available: \url{https://creativecommons.org/share-your-work/licensing-considerations/compatible-licenses/}

\bibitem{spdx3}
The Linux Foundation, ``SPDX Specification v3.0.1,'' 2024. [Online]. Available: \url{https://spdx.dev}

\bibitem{cyclonedx}
OWASP Foundation, ``CycloneDX Bill of Materials Standard, v1.6,'' Ecma International Standard ECMA-424, 2024. [Online]. Available: \url{https://cyclonedx.org}

\bibitem{li2025oss}
B. Li, C. Liu, L. Fan, S. Chen, Z. Zhang, and Z. Liu, ``Open Source, Hidden Costs: A Systematic Literature Review on OSS License Management,'' \textit{IEEE Trans. Softw. Eng.}, vol. 51, no. 9, pp. 2432--2454, Sep. 2025.

\end{thebibliography}
\end{document}